\documentclass[11pt]{article}
\usepackage{moriond,epsfig}

\def\be{\begin{equation}}
\def\ee{\end{equation}}
\def\bea{\begin{eqnarray}}
\def\eea{\end{eqnarray}}

\begin{document}
\vspace*{4cm}
\title{PHOTON PLUS JET CROSS SECTIONS AT THE TEVATRON}

\author{L. Sonnenschein}

\address{CERN and LPNHE Paris, Universit\'es Paris VI, VII}

\maketitle\abstracts{
Photon plus jet production has been studied by the D\O\ and CDF experiments 
in Run~II of the Fermilab Tevatron Collider at a center of mass energy of $\sqrt{s}=1.96$~TeV.
Measurements of the inclusive photon plus jet, di-photon and photon plus $b$ jet cross section
are presented. They are based on integrated luminosities between 0.2~fb$^{-1}$ and 1.1~fb$^{-1}$.
The results are compared to perturbative QCD calculations in various approximations.}

\section{Introduction}

Photons originating from the hard subprocess and produced during fragmentation 
contribute to photon cross sections in hadron-hadron collisions. 
The contribution of fragmentation photons can 
be significantly reduced by isolation requirements. Thus, isolated photon cross sections are sensitive to
the dynamics of the hard subprocess, to the strong coupling constant $\alpha_s$ and to the parton
distribution functions (PDF's) of the colliding hadrons.
In particular di-photon final states provide also a signature of many interesting physics processes,
such as $H\rightarrow\gamma\gamma$ production or large extra dimensions.

\section{Photon plus jet cross section}

D\O\ has measured the photon plus jet cross section\cite{pj} 
based on an integrated luminosity of 1.1~fb$^{-1}$.
Photon candidates are defined as clusters of electromagnetic (EM) calorimeter cells within a cone
of radius $R=0.2$ in the space of pseudorapidity $\eta$ and azimuthal angle $\phi$,
if more than 96\% of the detected energy is located in the EM layers of the calorimeter
and the probability for a track match is below 0.001.
As isolation criterion the transverse energy not associated to the photon in a cone of radius 
$R=0.4$ around the photon direction has to be less than 0.07 times the energy of the photon.
Backgrounds from cosmics and electrons from $W$ boson decays are vetoed by a missing transverse
energy requirement of $E_T\!\!\!\!\!\!\!/\;\, < 12.5~\mbox{GeV} + 0.36p_T^{\gamma}$.
Electromagnetic cluster and track information is fed into a neural network (NN) to
increase the photon purity further. 
Central photons ($|\eta|<1.0$) with a transverse momentum above 30~GeV are selected.
Jets are defined in the energy scheme by the Run~II midpoint cone 
algorithm\cite{r2} with a radius of $R=0.7$.
The jets with a transverse momentum above 15~GeV are selected in the central 
($|\eta^{\mbox{\scriptsize jet}}|<0.8$) or the forward ($1.5<|\eta^{\mbox{\scriptsize jet}}|<2.5$) 
region.
Finally, the photon and the leading hadronic jet have to be separated by 
$\Delta R(\gamma,\mbox{\scriptsize jet})>0.7$.

\begin{figure}[t]
\vspace*{-2.5ex}
\hspace*{2.0ex}
\psfig{figure=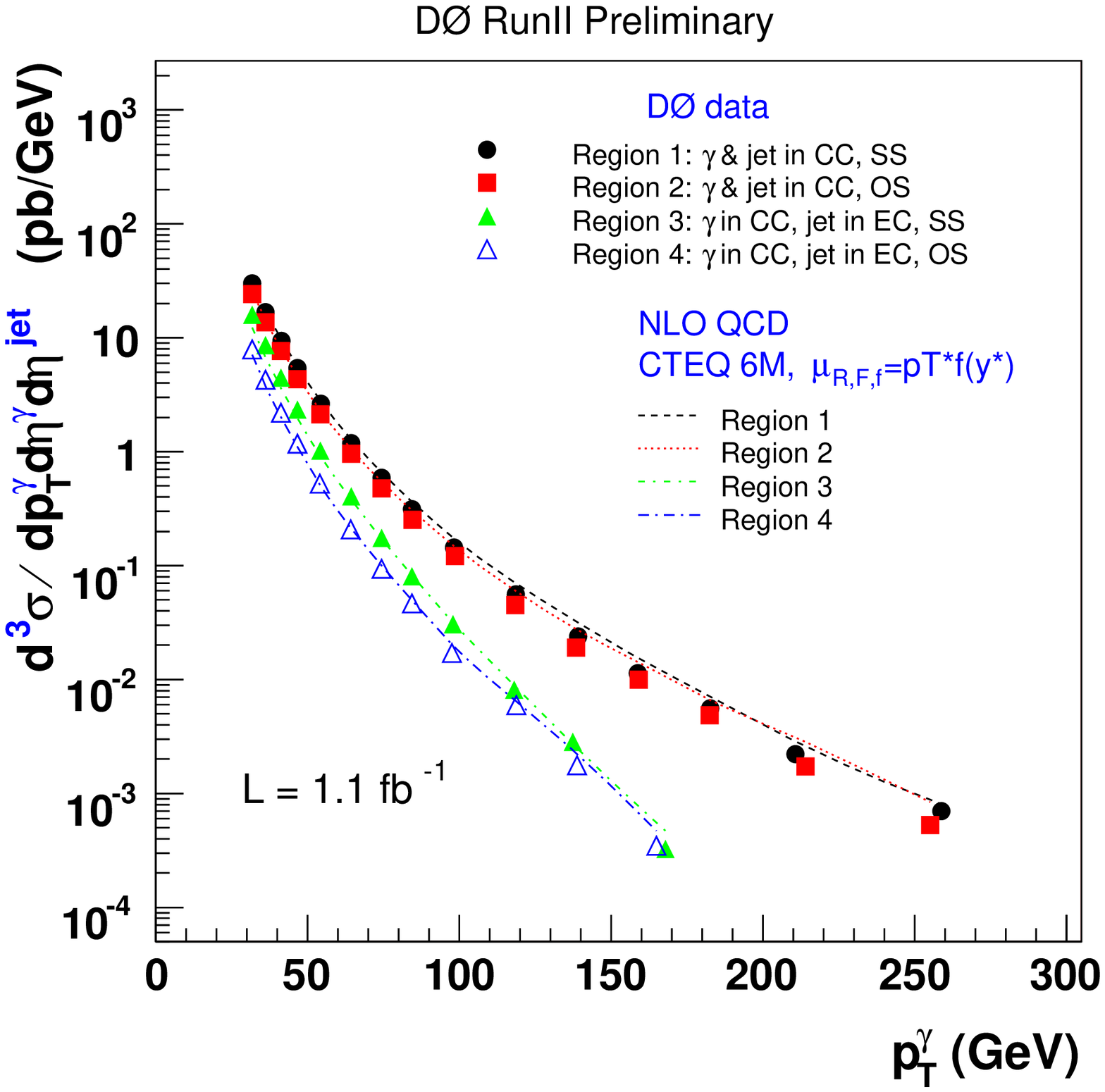,height=2.76in}
\psfig{figure=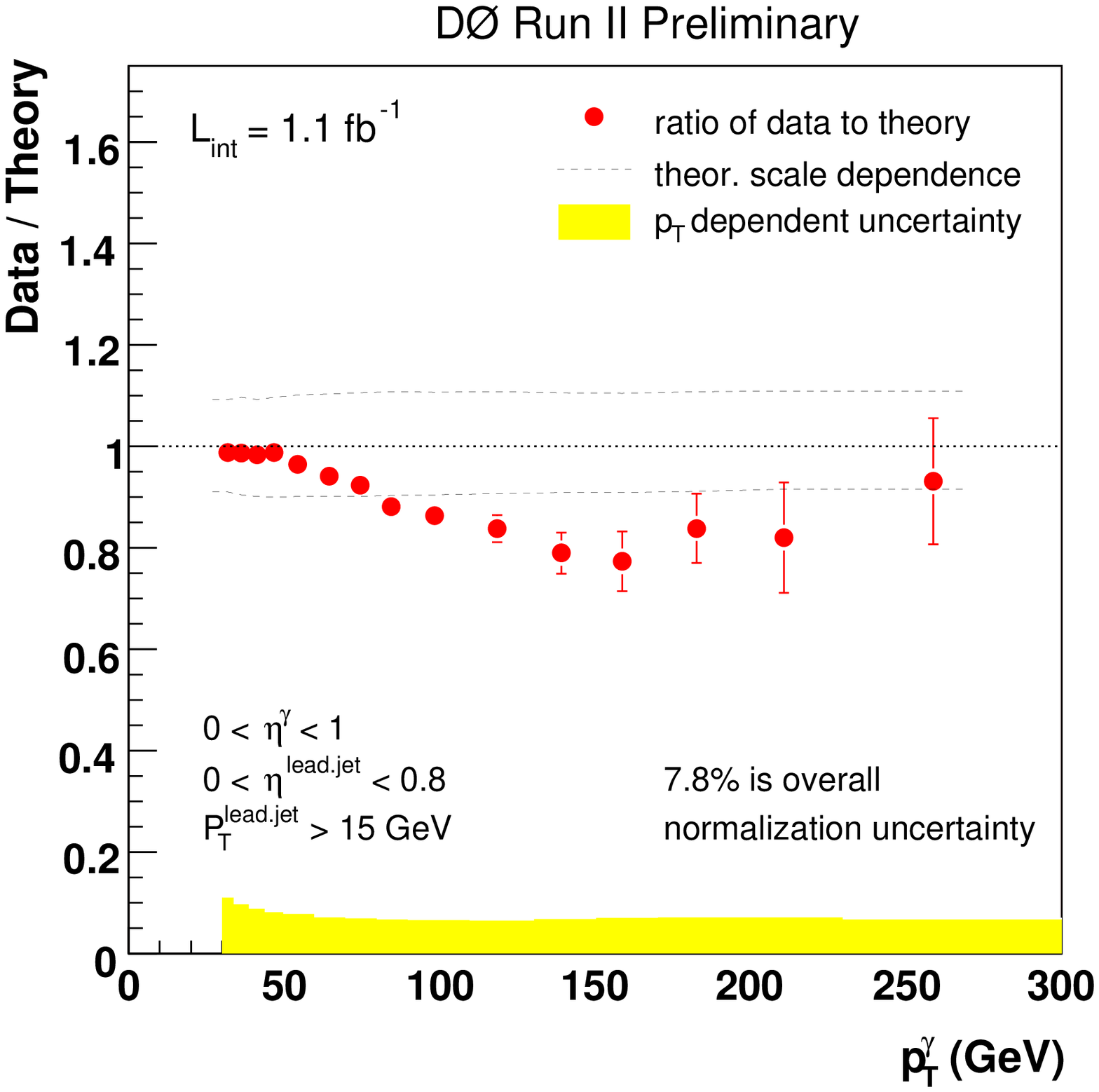,height=2.9in}
\hspace*{-4ex}
\vspace*{-3ex}
\caption{Differential photon plus jet cross section as a function of the photon transverse momentum
in four different kinematic regions (left),
the data/theory ratio in one kinematic region, (right). 
\label{fig:gammajet}}
\vspace*{-1ex}
\end{figure}

Isolated photons from photon plus jet production come predominantly unaltered 
from the hard subprocess. Thus, they provide a direct probe of the hard scattering dynamics.
At low transverse photon momentum the Compton scattering process $qg\rightarrow q\gamma$ dominates,
which can be exploited to probe the gluon density of the colliding hadrons at low $x$ 
where quarks are constrained by HERA data.
At higher transverse momenta the quark anti-quark annihilation process $q\bar{q}\rightarrow g\gamma$
becomes more important. The production of photons by quark fragmentation is highly suppressed
by the photon isolation.

The differential photon plus jet cross section is measured as a function of the transverse 
photon momentum. In fig. \ref{fig:gammajet} its $p_T^{\gamma}$-weighted average, obtained by a fit 
in each bin is plotted for four kinematic regions, 
distinguished by same side (SS) versus opposite side (OS) photon and jet and distinguished 
by central (CC) versus forward (EC) jet. 
In comparison the NLO pQCD prediction of JETPHOX\cite{px} is shown, making use of the 
CTEQ6.1M\cite{61} PDF's and BGF\cite{bg} fragmentation functions.
The scales have been chosen\cite{gf} to be $\mu_{R,F,f}=p_T^{\gamma}f(y^*)$ 
with $f(y^*)=\sqrt{(1+\exp(-2|y^*|))/2}$
and $y^*=0.5(\eta^{\gamma}-\eta^{\mbox{\scriptsize jet}})$.
The right plot shows the data/theory ratio in the first region
(same side photon and central jet). The scale variation by a factor of two up and down is not able
to describe the data at medium transverse photon momenta. Previous observations of UA2, CDF and D\O\
show a similar variation.
The systematics can be further reduced in considering the ratio of two regions. 
The ratios of region~1/region~3 and region~2/region~3 exhibit quantitative disagreement between 
data and theory.

\section{Di-photon cross sections}

Prompt di-photon production has been measured by CDF\cite{dp} using a data sample of 207~pb$^{-1}$.
Central photon candidates ($|\eta_{\gamma}|<0.9$) are selected based on lateral shower profile, 
preshower hit and no associated track requirements. 
As isolation criterion the transverse energy not associated to the photon in a cone of radius 
$R=0.4$ around the photon direction has not to exceed 1~GeV.

\begin{figure}[t]
\vspace*{-2.5ex}
\hspace*{-1ex}
\psfig{figure=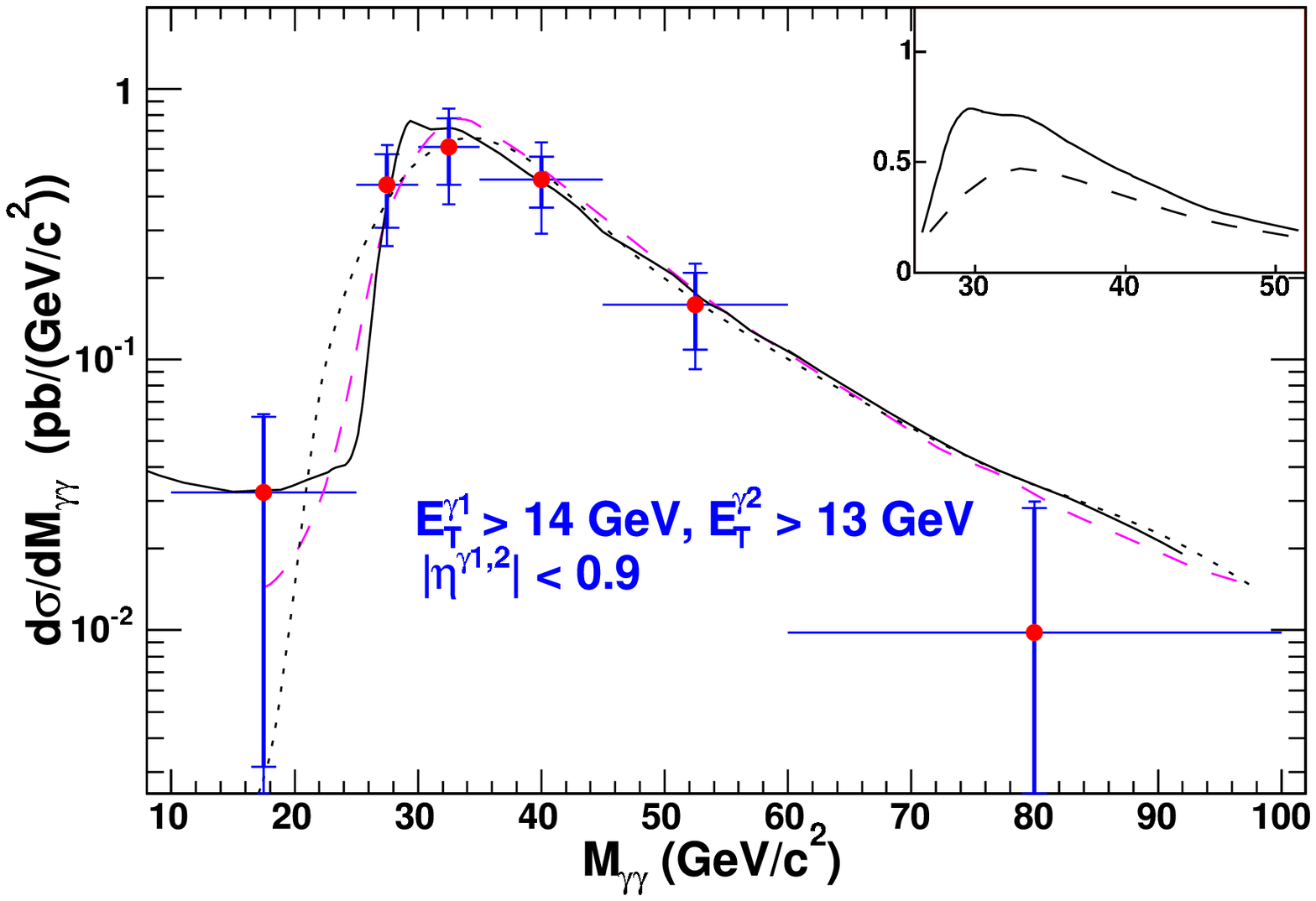,height=2.25in}
\hspace*{-3ex}
\psfig{figure=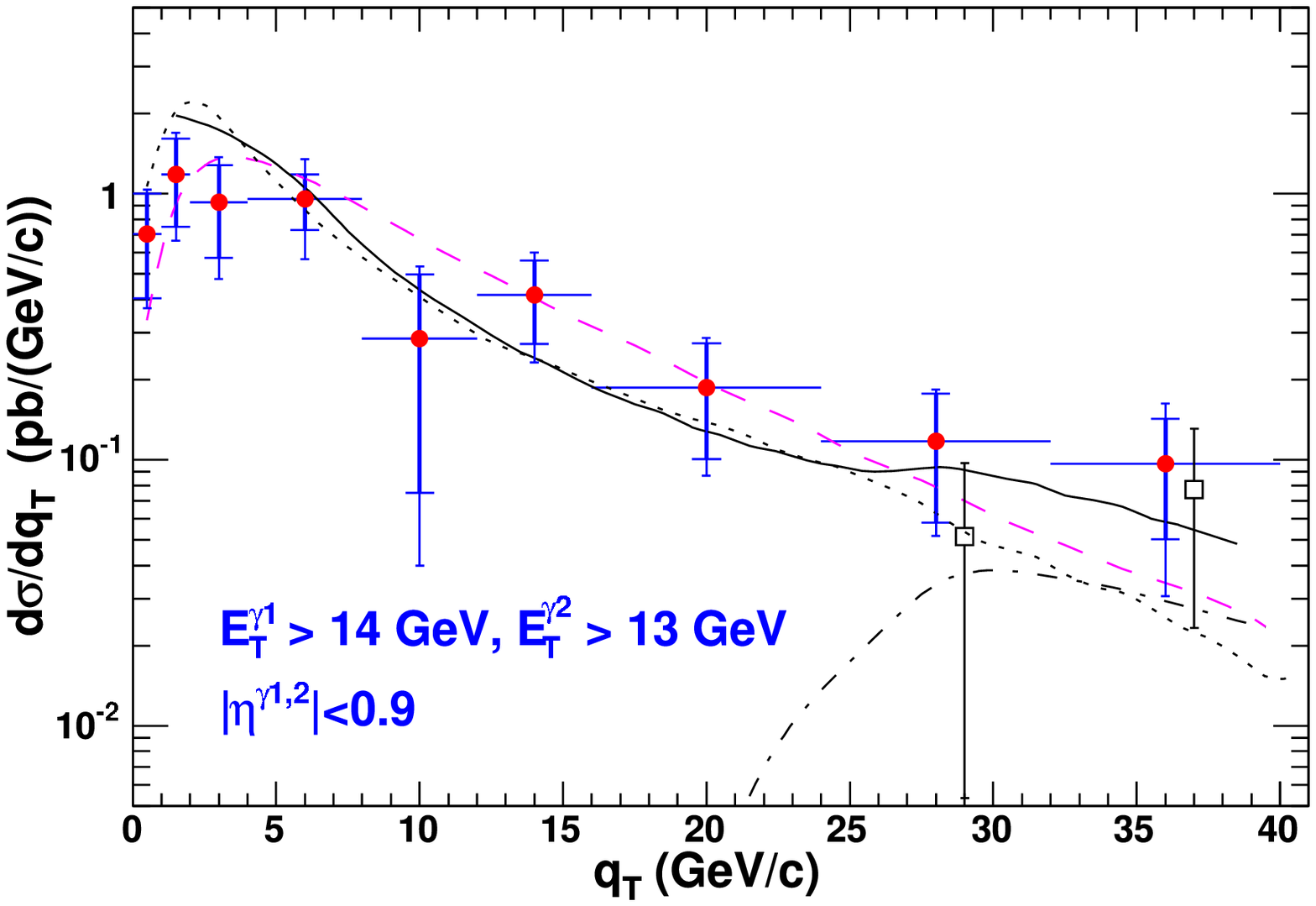,height=2.25in}
\vspace*{-5ex}
\caption{Differential di-photon cross section as a function of the invariant di-photon mass 
$m_{\gamma\gamma}$ (left)
and as a function of the transverse momentum $q_T$ of the di-photon system (right),
in comparison to the predictions of DIPHOX (solid line), ResBos (dashed line) and Pythia normalised to data (dotted line).
\label{fig:digamma}}
\vspace*{-1ex}
\end{figure}

Leading contributions to di-photon production come from quark-antiquark annihilation and gluon-gluon
scattering. 
A further contribution comes from quark fragmentation into a photon, 
which is suppressed by photon isolation but relevant in different phase space regions.

Fig. \ref{fig:digamma} shows the differential di-photon cross section as a function of the invariant 
di-photon mass (left) and the transverse momentum of the di-photon system (right)
in comparison to the theoretical predictions of DIPHOX\cite{tb}, ResBos\cite{rb} and LO Pythia\cite{py} 
increased by a factor of two to match the data.
While DIPHOX and ResBos contain the NLO contribution to prompt di-photon production, only DIPHOX
has a NLO fragmentation contribution for one or two photons, relevant at low mass, high $q_T$
and small $\Delta\phi_{\gamma\gamma}$. DIPHOX takes also into account the NNLO contribution of the
$gg\rightarrow\gamma\gamma$ process, indicated in the inset of the invariant mass distribution.
ResBos in turn has only a LO fragmentation contribution but takes into account resummed initial state
gluon radiation which is relevant at low $q_T$. Describing the
di-photon production in all regions of the phase space simultaneously requires a full NLO calculation,
taking ${\cal{O}}(\alpha_s^3)$ $gg\rightarrow\gamma\gamma$ corrections and resummed initial 
state gluon radiation into account.

CDF also searched for $gg\rightarrow \gamma\gamma$ with no other particles produced, 
i.e. exclusive $p\bar{p}\rightarrow p +\gamma\gamma + \bar{p}$. Two candidate events were found\cite{ta} 
consistent with theoretical expectations\cite{ak}.

\section{Photon plus $b$ jet cross sections}

Photon plus heavy flavour (HF) jet production is dominated by the process $g+$HF$\rightarrow\gamma+$HF.
This can be exploited to set constraints on the HF content of PDF's.

The photon plus $b$ jet production cross section has been measured by CDF\cite{pb} 
using an integrated luminosity of 340~pb$^{-1}$. 
A central photon ($|\eta_{\gamma}|<1.1$) with a transverse energy above 26~GeV
and a jet with a transverse energy above 20~GeV within $|\eta|<1.5$ and a secondary 
vertex are selected.
The invariant mass distribution of the secondary vertex of simulated bottom, charm and light quark jet 
contributions is fitted to the data in order to subtract the charm and light background.
This procedure is done separately in each bin of photon transverse energy to determine 
the differential photon plus $b$ jet cross section as a function of the photon transverse energy.
The data is in good agreement with the predictions of LO Pythia and Herwig\cite{HW}, making use of the
CTEQ5L\cite{CT} PDF's.

A similar CDF analysis\cite{pq} makes use of a dedicated photon plus displaced track trigger,
exploiting an integrated luminosity of 208~pb$^{-1}$. 
The transverse energy requirement of the central photon
($|\eta_{\gamma}|<1.1$) has been lowered to 12~GeV which extends the used phase space considerably.
As before, a jet with a transverse energy above 20~GeV within $|\eta|<1.5$ and an associated secondary 
vertex is required. The invariant mass distribution of the secondary vertex is fitted and the 
differential cross section as a function of photon transverse energy as well as the
differential cross section as a function of the jet transverse energy are determined.
The total cross section amounts to $90.5 \pm 6.0$(stat.) $^{+21.7}_{-15.4}$(sys.)~pb.
The LO prediction of Pythia with CTEQ5L PDF's slightly underestimates the cross section, 
whose accuracy is already systematics limited, dominated by the tracking efficiency uncertainty.

\section{Summary}
Inclusive photon plus jet, di-photon and photon plus $b$ jet cross sections have been measured
by the D\O\ and CDF experiments in $p\bar{p}$ collisions at a center of mass energy of $\sqrt{s}=1.96$~TeV.

The differential photon plus jet cross section 
has been measured in four different kinematic regions and their ratios. 
Deviations from the prediction of NLO QCD calculations are observed. 
The structure is similar to previous observations of UA2, CDF and D\O.

The di-photon invariant mass, transverse momentum and azimuthal opening angle
dependence of the di-photon cross section can not be described simultaneously by
a single theoretical prediction. In different regions of phase space, different contributions are
relevant. The different contributions consist of NLO corrections for the direct photon and 
fragmentation process, initial state soft gluon resummation and NNLO (${\cal{O}}(\alpha_s^3)$) 
corrections to the $gg\rightarrow\gamma\gamma$ process. A computation, combining all
separate contributions is needed to describe the di-photon production in all
phase space regions.
 
The photon plus $b$ jet cross section has been measured without and with a dedicated trigger,
based on a photon plus displaced track requirement. 
The latter analysis extends the phase space considerably in
lowering the photon transverse energy threshold from 25 to 12~GeV.
While LO predictions show good agreement with the cross section of the former analysis they 
slightly underestimate the cross section of the latter analysis.

\section*{Acknowledgements}
Many thanks to the stuff members at Fermilab, collaborating institutions and the QCD groups
of the D\O\ and CDF experiments.
Among others, this work has been supported by the Marie Curie Program,
in part under contract number MRTN-CT-2006-035606
and the HEPtools EU Marie Curie Research Training Network under contract number 
MRTN-CT-2006-035505.

\end{document}